\pdfoutput=1

\documentclass[12pt]{article}
\usepackage[colorlinks=true, urlcolor=blue, linkcolor=blue, citecolor=blue]{hyperref} %hyperlinks
\usepackage{amsmath,amsthm,amsfonts,amssymb,amscd} % all the usual ams symbols
\usepackage{bm}
\usepackage{subcaption} % allow subcaptions!
\usepackage{graphicx} % graphics
\usepackage[font={small}]{caption} % small text in captions
\usepackage{booktabs} % fancy tables
\usepackage[authoryear]{natbib} % citep
\usepackage{setspace} % spacing
\usepackage{authblk}
\usepackage{dsfont}
\usepackage{float}
\usepackage{xcolor}
\usepackage{parskip}
\usepackage{indentfirst}
\usepackage[left= 1in,right=1in,top=1in,bottom=1in]{geometry}
\usepackage{fancyhdr}
\newcommand{\tp}{\mathsf{T}}
\newcommand{\caln}{\mbox{${\cal N}$}}

\newcommand{\alp}{\mbox{\boldmath $\alpha$}}
\newcommand{\bet}{\mbox{\boldmath $\beta$}}

\newcommand{\eps}{\mbox{\boldmath $\epsilon$}}

\newcommand{\U}{\mbox{\boldmath $U$}}
\newcommand{\V}{\mbox{\boldmath $V$}}
\newcommand{\Z}{\mbox{\boldmath $Z$}}
\newcommand{\R}{\mbox{\boldmath $R$}}
\newcommand{\X}{\mbox{\boldmath $X$}}
\newcommand{\E}{\mbox{\boldmath $E$}}
\newcommand{\I}{\mbox{\boldmath $I$}}

\newcommand{\Y}{\mbox{\boldmath $Y$}}
\newcommand{\B}{\mbox{\boldmath $B$}}

\newcommand{\bu}{\mbox{\boldmath $u$}}
\newcommand{\bx}{\mbox{\boldmath $x$}}
\newcommand{\bv}{\mbox{\boldmath $v$}}
\newcommand{\bs}{\mbox{\boldmath $s$}}
\newcommand{\br}{\mbox{\boldmath $r$}}
\newcommand{\bze}{\mbox{\boldmath $0$}}
\newcommand{\bone}{\mbox{\boldmath $1$}}

\newcommand{\hbv}{\mbox{$\hat{\bv}$}}
\newcommand{\hbu}{\mbox{$\hat{\bu}$}}

\title{Probabilistic Predictive Principal Component Analysis for Spatially-Misaligned and High-Dimensional Air Pollution Data with Missing Observations}

\author[1]{Phuong T. Vu}
\author[2]{Timothy V. Larson}
\author[1]{Adam A. Szpiro}
\affil[1]{Department of Biostatistics, University of Washington}
\affil[2]{Department of Civil \& Environmental Engineering, University of Washington}

\footnotetext{\noindent Corresponding author: Phuong T. Vu, Department of Biostatistics, University of Washington, F-600, Health Sciences Building, Box 357232, Seattle, WA 98195-7232. Email: phuongvu@uw.edu.}

\date{August 22, 2019}

\begin{document}

\maketitle
\doublespacing

\begin{abstract}
Accurate predictions of pollutant concentrations at new locations are often of interest in air pollution studies on fine particulate matters (PM$_{2.5}$), in which data is usually not measured at all study locations. PM$_{2.5}$ is also a mixture of many different chemical components. Principal component analysis (PCA) can be incorporated to obtain lower-dimensional representative scores of such multi-pollutant data. Spatial prediction can then be used to estimate these scores at new locations. Recently developed predictive PCA modifies the traditional PCA algorithm to obtain scores with spatial structures that can be well predicted at unmeasured locations. However, these approaches require complete data, whereas multi-pollutant data tends to have complex missing patterns in practice. We propose probabilistic versions of predictive PCA which allow for flexible model-based imputation that can account for spatial information and subsequently improve the overall predictive performance.

\begin{center}{\small \textbf{Keywords:} air pollution, multi-pollutant analysis, missing data, dimension reduction}\end{center}
\end{abstract}

\newpage
\doublespacing

\section{Introduction \label{sec-intro}}

In recent years, there has been a growing interest in studying the role and health impact of PM$_{2.5}$, which is fine particulate matter with aerodynamic diameter less than 2.5 $\mu$m \citep{brook2004air}. PM$_{2.5}$ is a complex mixture of many components, and its chemical profile may vary drastically across time and space \citep{brook2004air, bell2007spatial, dominici2010protecting}. Obtaining a lower-dimensional representation of PM$_{2.5}$ multi-pollutant data is often necessary, as including many correlated pollutants in a statistical model is problematic. Principal component analysis (PCA) \citep{jolliffe1986principal} is an unsupervised dimension reduction technique that has gained popularity in multi-pollutant analysis \citep{dominici2003health}. 

Examples of environmental studies utilizing PM$_{2.5}$ data include studies on the associations between various health outcomes and long-term \citep{pope2002lung, kunzli2005ambient, miller2007long, chan2015long, kaufman2016association} or short-term \citep{gold2000ambient, tolbert2007multipollutant, pascal2014short, achilleos2017acute, hsu2017ambient, tian2017addressing} exposures to PM$_{2.5}$. Many studies have suggested that the associations between PM$_{2.5}$ total mass and various health outcomes can be modified by some specific constituents or the overall chemical composition \citep{franklin2008role, bell2009hospital, krall2013short, zanobetti2014health, dai2014associations, kioumourtzoglou2015pm2, wang2017long, keller2018pollutant}.

In the United States, PM$_{2.5}$ studies often rely on data collected from regulatory monitoring networks managed by the Environmental Protection Agency (EPA). Unfortunately, for many pollution-health association studies, these fixed monitoring sites are usually not at the same locations where health outcomes are available. Such \textit{spatial misalignment} motivates an exposure modeling stage in which a spatial prediction model, such as land-use regression or universal kriging, is often used to estimate the exposure at unmeasured locations where pollutant data is not observed \citep{brauer2003estimating, kunzli2005ambient, crouse2010postmenopausal, bergen2013national, chan2015long}. 

Derivation of a lower-dimensional representation of PM$_{2.5}$ multivariate data prior to making these spatial predictions is necessary, as predicting chemically and spatially correlated pollutant surfaces is challenging and intractable in most cases. As PCA is capable of performing dimension reduction without meddling with the health outcomes, it can be easily integrated in the analysis of spatially-misaligned data. Using PCA, a lower-dimensional scores of the multi-pollutant data at monitoring locations can be obtained. These monitoring scores, along with geographic covariates, can then be used in a spatial prediction model to estimate the corresponding scores at unmeasured locations. However, PCA does not account for exogenous geographic information and spatial correlations across neighboring locations. Hence, PCA may produce scores that summarize the monitoring data well but are difficult to be predicted at unmeasured locations. A spatially predictive PCA algorithm \citep{jandarov2017novel} was developed to mitigate this issue by producing scores with spatial patterns that can be subsequently predicted well at new locations. 

An additional challenge arises in practice where there is often a large amount of missing data, especially for multi-pollutant monitoring data. For example, not all PM$_{2.5}$ components are measured at all monitoring sites, either due to environmental considerations, logistic constraints or lack of resources. The missing patterns can sometimes be complex or spatially informative. Neither traditional PCA nor predictive PCA is well-equipped to deal with missing data, and thus a separate imputation step is required prior to dimension reduction. Existing non-parametric imputation schemes, ranging from simple mean imputation to sophisticated matrix completion, do not account for external spatial information. They may therefore distort the underlying spatial structure in the original data even before the dimension reduction stage, and thus negatively impact the predictive performance in the final stage. 

In this paper, our goal is to enhance the dimension reduction procedure under the presence of missing data by proposing a probabilistic framework in place of the deterministic algorithm of predictive PCA. Similar to \cite{jandarov2017novel}, our methods seek to produce principal components that can be well predicted at new locations. The added probabilistic assumptions allow for flexible model-based imputation that takes into account the embedded geographic and spatial information, and thus eliminates the need for a preprocessing stage with non-parametric imputation.

\section{Motivating example \label{sec-motivating}}

To illustrate the merit of our proposed methods, we use data collected nationally by the Air Quality System (AQS) network of monitors managed by the EPA. Measurements of annually averaged PM$_{2.5}$ total mass and its components are only collected at a few subnetworks of AQS. For consistency with previous related work \citep{keller2017covariate, jandarov2017novel}, we choose to use the 2010 data from the Chemical Speciation Network (CSN), of which monitoring sites are located strategically in various urban areas. Data is available for 21 components of PM$_{2.5}$: elemental carbon (EC), organic carbon (OC), sulfate ion (SO$^{2-}_4$), nitrate ion (NO$^{-}_3$), aluminum (Al), arsenic (As), bromine (Br), cadmium (Cd), calcium (Ca), chromium (Cr), copper (Cu), iron (Fe), potassium (K), magnesium (MN), sodium (Na), sulfur (S), silicon (Si), selenium (Se), nickel (Ni), vanadium (V), and zinc (Zn). 

Geographic covariates are obtained for all available sites through the Exposure Assessment Core Database by the MESA Air team at the University of Washington. Data on roughly 600 Geographic Information System (GIS) covariates are available, including distances from roads, distances from major pollution sources, land-use information, vegetation indices, etc. The specific sources and attributions of these geographic covariates are carefully described in \cite{bergen2013national}. 

Data for 2010 is available for 221 CSN sites, with only 130 of those sites having complete data on all 21 components. Overall the amount of missing data in 2010 is roughly 32.1\%. Not only do we compare the predictive performances following the application of different PCA methods, but we also examine how different the chemical profiles are when considering only complete sites versus all available data. The data processing, analysis procedures, and results are discussed in Section \ref{sec-application}.

\section{Review of PCA and predictive PCA \label{sec-review}}

We denote $\X \in \mathds{R}^{n \times p}$ as the exposure data with $p$ pollutants observed at $n$ monitoring sites with spatial coordinates $\bs_1, ..., \bs_n$. The exposure data $\X$ may contain missing elements as some pollutants are not measured at all monitoring site. Let $\br_i$ be a vector of $k$ geographic covariates pertaining to the $i$-th monitoring sites. Variables corresponding to locations where exposure data is of interest but not measured are distinguished by an asterisk, i.e. $n^*, \X^*, \bs^*_1, ..., \bs^*_{n^*}, \br^*_1, ..., \br^*_{n^*}$. 

The data of interest, $\X^*$, is high-dimensional but inaccessible. If $\X^*$ were observed, dimension reduction could be applied directly to obtain a lower-dimensional representation $\U^* \in \mathds{R}^{n^* \times q}$ where $q < p$. Because of spatial misalignment, a spatial prediction model is required to estimate the unobserved exposures. Modeling highly correlated surfaces is challenging and inefficient given the final aim of recovering only the lower-dimensional $\U^*$. Thus, a sensible modeling procedure under the presence of spatially misaligned multi-pollutant data with missing observations may consist of several steps: (1) imputation for missing data, (2) dimension reduction to derive scores at monitoring sites, and (3) spatial prediction to estimate corresponding scores at new locations. In this paper, we focus on dimension reduction using PCA, an unsupervised technique that is suitable for handling spatially-misaligned data. 

Traditional PCA provides a mapping from the original $p$-dimensional exposure surface to a corresponding $q$-dimensional representation where $\X \approx \U\V^\tp$ for $q<p$. We refer to the orthogonal columns of $\V \in \mathds{R}^{p\times q}$ as the loadings or principal directions. The columns of $\U \in \mathds{R}^{n\times q}$, $\{\bu_1, ..., \bu_q \}$, are the principal component (PC) scores. These PC scores can be thought of as linear combinations of the original features of $\X$. These newly transformed variables are considered uncorrelated due to orthogonality of the loadings, which is an attractive feature of PCA. The PCA algorithm is also optimal in the sense that the derived PC scores are conveniently ordered by the amount of variability explained in $\X$. 

While PCA provides a unique solution in the reduced dimensions, the algorithm can be reformulated into a series of biconvex optimization problems, in which the loading and corresponding score of each PC can be solved in an iterative fashion \citep{shen2008sparse}, 
$$\min_{\bu, \bv} \Big\Vert \X - \bu\bv^\tp  \Big\Vert^2_F \hspace{0.25cm} \text{s.t.} \hspace{0.1cm} \Vert \bv \Vert_2 = 1.$$
Utilizing such optimization framework, \cite{jandarov2017novel} develop a spatially predictive PCA algorithm (PredPCA hereafter) by directly incorporating spatial information in the objective function: 
$$\min_{\alp, \bv}   \bigg\Vert \X - \left( \frac{\Z \alp}{\Vert \Z \alp \Vert_2} \right) \bv^\tp \bigg\Vert^2_F ,$$
where $\Z = \begin{bmatrix}
\R & \tilde{\R}
\end{bmatrix}$, in which $\R \in \mathds{R}^{n \times k}$ contains $k$ GIS covariates, and $\tilde{\R} \in \mathds{R}^{n \times \tilde{k}}$ contains $\tilde{k}$ thin-plate spline basis functions. The induced PC score, $\Z \alp/ \Vert \Z \alp \Vert_2$, is constrained to have an underlying smooth spatial structure guided by geographic and spatial information encoded in $\Z$. An advantage of PredPCA over PCA is the capability to identify principal directions that lead to spatially predictable PC scores at unmeasured locations. Recent work by \cite{bose2018adaptive} further improves PredPCA by adaptively selecting information to be included in $\Z$ for each PC.  

When monitoring data is incomplete, simply omitting locations with missing data may reduce the usable sample size substantially; thus,  imputation is often required prior to dimension reduction. Non-parametric techniques, ranging from mean imputation to matrix completions, are based on observed pollutant values but not additional spatial information. When the missingness is spatially informative, such imputation schemes may heavily bias the results of these algorithms. 

In the next section, we propose a probabilistic framework that aims to derive spatially predictive PC scores, with the ability to handle incomplete monitoring data and induce flexible model-based imputation that accounts for spatial and geographic information.

\section{Probabilistic predictive PCA \label{sec-proposed}}

\subsection{Probabilistic formulation with a latent variable model: the Krige algorithm}

\cite{tipping1999probabilistic} proposed a probabilistic formulation of PCA based on a Gaussian latent variable model. Their model assumes $\X = \bu \bv^\tp + \E$, where $\bu \sim \mathcal{N}(\bze, \I_n)$, $\bv \in \mathds{R}^p$, $\Vert \bv \Vert_2 = 1$, and the elements of $\E$ are independently and identically distributed (i.i.d.) with mean zero and variance $\gamma^2$. We extend this framework by directly imposing a spatial mean and covariance structure on the latent variable space. That is, given a desired number of PCs, $q$, our model assumes
\begin{align*}
\X &= \sum^q_{l = 1} \left( \bu_l \bv_l^\tp + \E_l  \right),\\
\bu_l &= \R \bet_l + \boldsymbol{\eta}_l,
\end{align*}
where $\bet_l \in \mathds{R}^k$ includes the coefficients corresponding to the geographic covariates in $\R$, while $\boldsymbol{\eta}_l \in \mathds{R}^n$ has zero mean and spatial covariance $\Sigma(\boldsymbol{\xi}_l)$, with $\boldsymbol{\xi}_l$ denoting the spatial covariance parameters of the latent space. We use similar constraint $\Vert \bv_l \Vert_2 = 1$, and assume that $\Sigma(\boldsymbol{\xi}_l)$ has no nugget effect. The latent score $\bu_l$ is stochastic with a full spatial distribution. 

Let $\Theta_l$ be the collection of the model parameters, $\{\bv_l, \bet_l, \gamma_l^2, \boldsymbol{\xi}_l\}$, corresponding to the $l$-th PC. When the monitoring data is complete, estimate of the first loading, $\hbv_1$, can be obtained using the original data matrix $\X$. The corresponding score $\hbu_1$ at monitoring locations can then be calculated by projecting $\X$ onto the direction of $\hbv_1$. In later steps, $\Theta_l$ can be estimated using $\X_l = \X_{l-1} - \hbu_{l-1} \hbv_{l-1}^\tp$, where $\X_1 = \X$. The PC score $\hbu_l$ can then be derived by projecting $\X_l$ onto $\hbv_l$. Note that we use projection of the data matrix to obtain the PC score in each step instead of using model estimate of the latent mean $\R\bet_l$. When some elements of $\X$ are missing, estimation of $\Theta_l$ is based only on the observed elements of $\X_l$. Estimated PC score $\hbu_l$ can then be made by projecting the model-based imputed exposure data onto the direction of $\hbv_l$. 

Our approach to estimate $\Theta_l$ in each step is similar to the EM algorithm employed by \cite{tipping1999probabilistic}. We consider the latent variable $\bu_l$ to be the ``missing" portion, and thus the ``complete" data consists of the observed $\X_l$ and the latent variable $\bu_l$. The goal is then to maximize the joint likelihood of $\X_l$ and $\bu_l$. The mathematical details and algorithms for both complete and missing monitoring data are described in the Supplemental Materials. We refer to this framework as the probabilistic predictive PCA, or ProPrPCA, hereafter. Specifically, we call this algorithm ProPrPCA-Krige due to the kriging formulation in the model assumptions.

Our ProPrPCA-Krige model is closely related to the SupSVD model recently proposed by \cite{li2016supervised}. The SupSVD model is expressed as $\X = \U \V^\tp + \E$ where $\U = \Y \B + \mathbf{F}$. Here $\U$ is a the latent score matrix, $\V$ is a full-rank loading matrix, $\mathbf{F}$ and $\E$ are error matrices. \cite{li2016supervised} also propose an EM approach to estimate the model parameters. The ProPrPCA-Krige model is also related to the envelope model proposed in \cite{cook2010envelope}, which is a more general version compared to SupSVD. As discussed in \cite{li2016supervised}, the SupSVD model attempts to extract a low-rank representation of the original data based on some auxiliary data, while the envelope model aims to reduce variation in regression coefficient estimation. We note that our model is motivated by spatial misalignment where data are not observed at cohort locations, but some geographic information is available. The end goal is also different from the SupSVD and envelope models, as we seek to accurately predict a low-rank representation of the data at unmeasured locations. Thus, our model is designed such that patterns of available covariates and spatial structure are properly induced in the latent scores at locations where we have data, so that we can easily predict them at new locations. An additional contribution is that we develop EM algorithms for parameter estimation for both complete and missing data scenarios.

\subsection{Probabilistic formulation with thin-plate spline basis: the Spline algorithm}

While the ProPrPCA-Krige algorithm is cohesive with a prediction stage using universal kriging, the parameter estimation appears to be computational burdensome. In general, the EM algorithm is often computationally expensive and convergence is not always guaranteed. We propose a more simplified version of ProPrPCA, %Inspired by the objective function of PredPCA, w
$$\X = \sum^q_{l=1}\left( (\Z\bet_l)\bv_l^\tp + \E_l  \right),$$
where $\Z$ contains thin-plate spline functions similar to PredPCA. Compared to the ProPrPCA-Krige model, the latent score $\bu_l$ no longer has a stochastic component. Instead, $\bu_l$ is now a smooth structure enriched with spatial patterns included in $\Z$. 

The overall procedure to obtain PC scores is similar to the Krige algorithm. The algorithm with complete monitoring data is shown in Table \ref{tab-spline-alg}. When some elements of $\X_l$ are missing, estimation of $\hat{\Theta}_l = \{\bv_l, \bet_l, \gamma_l^2\}$ is based on the observed elements of $\X_l$, and estimated PC score $\hat{\bu}_l$ can be derived by projecting the model-based imputed exposure matrix onto the direction of $\hbv_l$. When the monitoring data is complete, the algorithm for parameter estimation at each step is straightforward. The mathematical derivations and the algorithm for missing data are described in the Supplemental Materials. We refer to this model as ProPrPCA-Spline due to the use of thin-plate spline basis functions.

\section{Simulations \label{sec-simulation}}

We conduct two sets of simulations to compare the different PCA approaches. The first set involves a low-dimensional setting with three-pollutant exposure surfaces. The second set illustrates a higher-dimensional setting with 15 generated pollutant surfaces. In both cases, the multi-pollutant data is generated on a $100 \times 100$ grid ($N = 10,000) $. 

In each simulation, we randomly choose 400 training locations and 100 testing locations. We then apply the four competing methods (PCA, PredPCA, ProPrPCA-Krige, and ProPrPCA-Spline) to the training data, $\X^{train}$, to obtain the corresponding loading $\hbv_l^{train}$ and score $\hbu_l^{train}$, for $l = 1, ..., q$ where $q$ is a desired number of PCs. We then use $\hbu_l^{train}$ and relevant covariate information to obtain $\hat{\bu}_l^{test}$, predicted scores at testing locations, in a universal kriging model with an exponential covariance assumption. Finally, we compare the predicted scores to the known scores, $\bu_l^{test}$, which is defined by projecting $\X^{test}$ onto the direction of $\hbv_l^{train}$.

We also consider various scenarios in which some training data is missing. These scenarios include missing completely at random (MCAR) , with 30\%, 35\%, and 40\% of missing data, and missing at random (MAR), in which the missing patterns are associated with the generated spatial covariates. When there is missing data, we apply low-rank matrix completion (LRMC) via the SoftImpute algorithm \citep{mazumder2010spectral} to fill in the missing entries prior to PCA and PredPCA. 

There are several metrics to evaluate the predictive performance. The metric of interest is the prediction R$^2$ adapted from \cite{szpiro2011does}, which reflects the correlation between $\hbu_l^{test}$ and $\bu_l^{test}$. We also look at the reconstruction error (RE), defined as $\Vert \X^{test} - \hat{\X}^{test} \Vert_F$ where $\hat{\X}^{test} = \hat{\U}^{test} (\hat{\V}^{train})^{\tp}$, $\hat{\U}^{test} = \begin{bmatrix}
\hbu_1^{test} & ... & \hbu_q^{test}
\end{bmatrix}$, and $\hat{\V}^{train} = \begin{bmatrix}
\hbv_1^{train} & ... & \hbv_q^{train}
\end{bmatrix}$. 

\subsection{Three-dimensional exposure surfaces}

We simulate three-dimensional surfaces with $\{\bx_1, \bx_2, \bx_3\}$, and three independent covariates $\{\br_1, \br_2, \br_3\}$. Only $\br_1 \sim \caln(\bze, \I_N)$ is ``observed" and thus used in the universal kriging model. Both $\br_2 \sim \caln(\bze, \I_N)$ and $\br_3 \sim \caln(\bze, \I_N)$ are unobserved and primarily used to induce correlations across $\{\bx_1, \bx_2, \bx_3\}$. We generate data such that $\bx_1 = 4\br_1 + 2\br_3 + \eps_1$, $\bx_2 = 3\br_2 + \eps_2$, and $\bx_3 = 2\br_1 + 4\br_2 + \eps_3$, where $\eps_1, \eps_2, \eps_3 \sim \caln(\bze, \Sigma)$, where $\Sigma$ has an exponential structure with partial sill $\sigma^2 = 3.5^2$, nugget $\tau^2 = 1$, and range $\phi = 50$. Under this setting, only $\bx_1$ and $\bx_3$ are predictable by $\br_1$. While not dependent on $\br_1$, $\bx_2$ is moderately correlated with $\bx_3$ via $\br_2$. We also generate a second set of data in which the errors $\eps_1, \eps_2, \eps_3 \sim \caln(\bze, \bone)$ . For MAR scenarios, $\bx_1$ is missing at training locations where $\br_1$ values are larger than its 80th sample percentile, while $\bx_2$ and $\bx_3$ have 20\% MCAR. We look at the first PC for these simulations, i.e. $q = 1$.

\begin{figure}[H]
	\centering
	\includegraphics[width=6.5in]{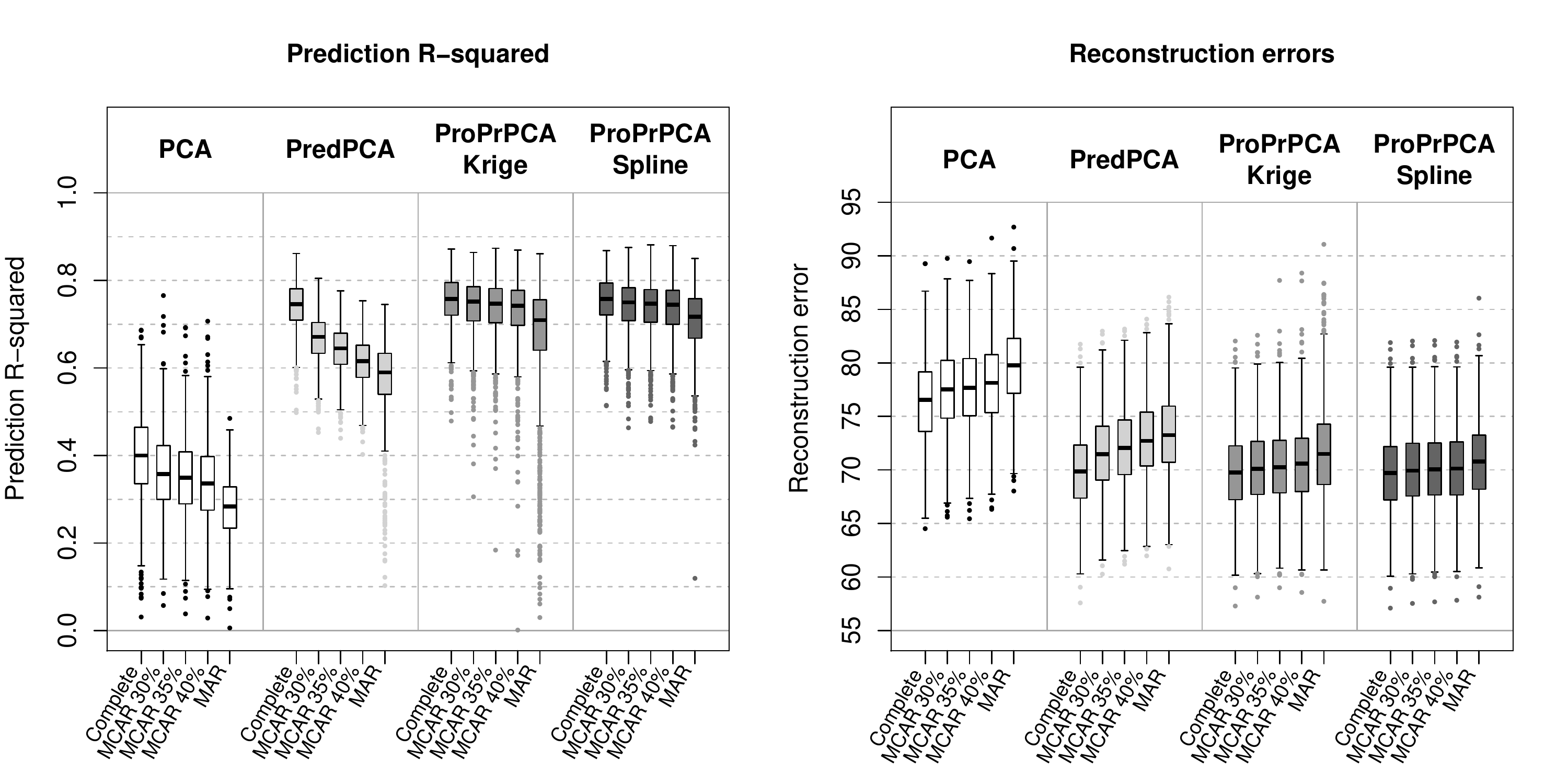}
	\caption{\doublespacing Prediction R$^2$'s and reconstruction errors across 1,000 replications with three-dimensional surface generated with spatially correlated noises. Under missing data scenarios, LRMC is used prior to the application of either PCA or PredPCA.}
	\label{fig-toy-scen2}
\end{figure}

Figure \ref{fig-toy-scen2} shows the prediction R$^2$'s and REs across 1,000 simulations for data generated with spatially correlated noise. Table \ref{tab-loading-scen2} displays the means and standard deviations of the estimated loadings from each method when the training data is complete. The principal direction produced by PCA is loaded heavily on $\bx_3$ and only moderately on both $\bx_1$ and $\bx_2$. This leads to poor predictive performance for PCA (median R$^2 = 0.40$). Meanwhile, loadings from the other three methods put the most weight on $\bx_1$ and some on $\bx_3$, thus they have higher prediction R$^2$'s (median R$^2$'s are about 0.75) and lower REs. 

Under MCAR scenarios, prediction R$^2$'s substantially decrease and REs increase for both PCA and PredPCA as the amount of missing data increases. Median R$^2$ of PredPCA drops to as low as 0.64 when training data is 35\% MCAR. On the other hand, there are only some subtle reductions in the predictive performances of both ProPrPCA approaches. Under MAR, the performances of both PCA and PredPCA are significantly worse. While ProPrPCA-Krige performs better than PredPCA on average, the variability in performance is high across simulations. Despite not achieving the same level as when the data is complete, ProPrPCA-Spline has the highest predictive performance among the four competing methods. 

\begin{figure}[H]
	\centering
	\includegraphics[width=6.5in]{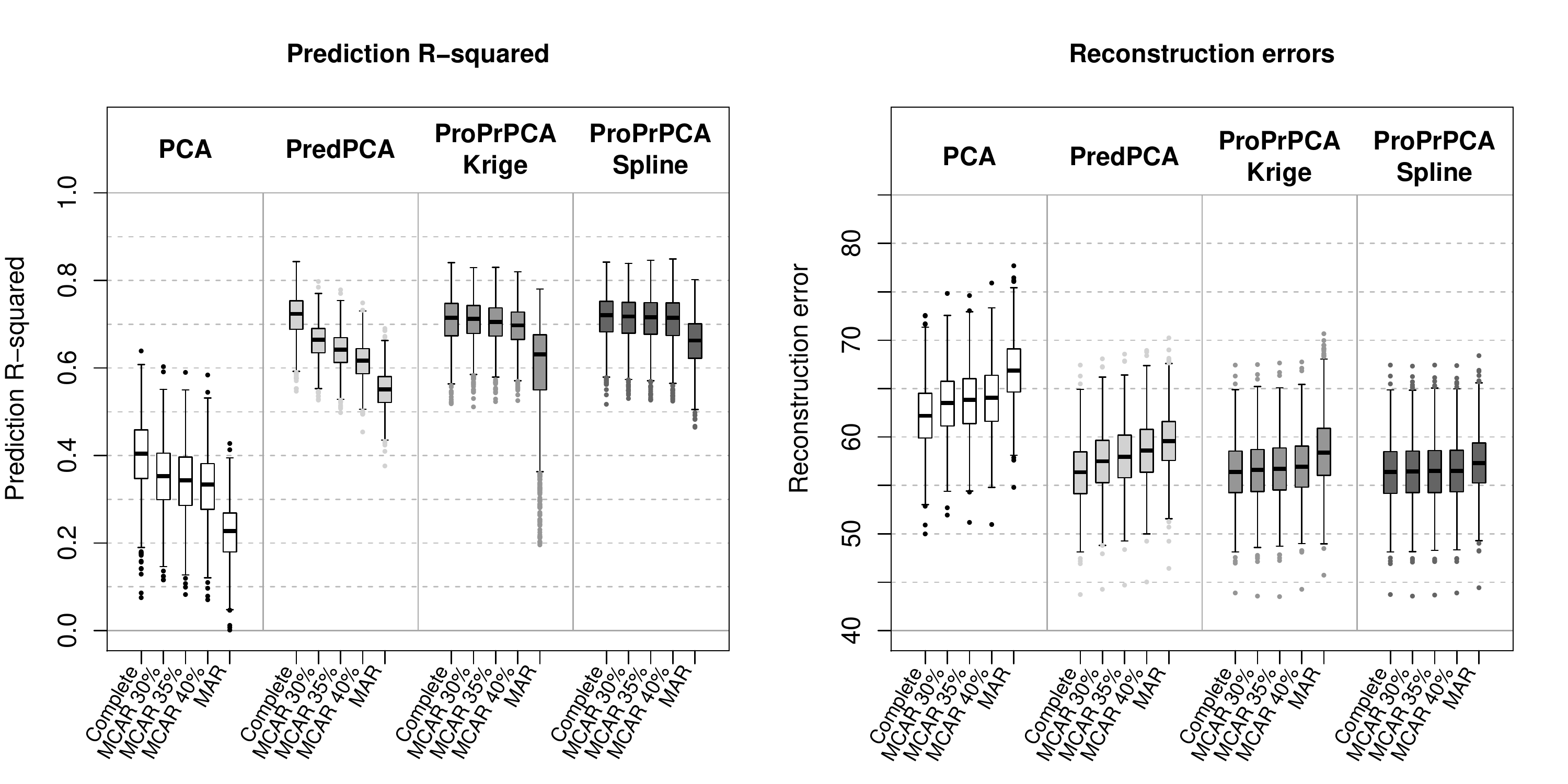}
	\caption{\doublespacing Prediction R$^2$'s and reconstruction errors across 1,000 replications with three-dimensional surface generated with independent noises. Under missing data scenarios, LRMC is used prior to the application of either PCA or PredPCA.}
	\label{fig-toy-scen1}
\end{figure}

Table \ref{tab-loading-scen1} shows the estimated loadings with complete data, while Figure \ref{fig-toy-scen1} shows the prediction R$^2$'s and REs across 1,000 simulations for data generated with independent noise. Similar trends, where ProPrPCA outperforms the rest when missing data is more severe, are also observed in this set of generated data.

\subsection{High-dimensional exposure surfaces}

We also demonstrate the performance of ProPrPCA algorithms via simulations with 15 generated pollutants. The full setup is described in the Supplemental Materials. Overall, the high-dimensional exposure surfaces are generated from three underlying scores, $\bu_1$, $\bu_2$, and $\bu_3$. The data generating mechanism is such that $\bu_1$ is the most spatially predictable, $\bu_2$ is moderately predictable, and $\bu_3$ is not predictable by any covariates used in the universal kriging model. The loadings used to generate the data are sparse, in order to clearly identify the behaviors of the PCA methods. That is, the first five pollutants, $(\bx_1, \bx_2, \bx_3, \bx_4, \bx_5)$, are generated from $\bu_1$. Meanwhile, $(\bx_6, \bx_7, \bx_8, \bx_9, \bx_{10})$ are generated from $\bu_2$, and $(\bx_{11}, \bx_{12}, \bx_{13}, \bx_{14}, \bx_{15})$ are generated from $\bu_3$. For MAR scenario, we induce a mild spatial pattern in the missing data for the first five pollutants. {\color{black}In these simulations, we evaluate the predictive performance based on two PCs, i.e. $q = 2$.}

We create two scenarios: scenario 1 with $Var(\bu_1) = 10$, $Var(\bu_2) = 7.5$, and $Var(\bu_3) = 5$, and scenario 2 with $Var(\bu_3) = 10$, $Var(\bu_1) = 7.5$, and $Var(\bu_2) = 5$. In scenario 1, where the order of variance contribution is the same as the order of spatial predictability, we expect all methods to identify {\color{black}linear combinations of $\bu_1$ and $\bu_2$ as the first two PCs} when training data is complete. In scenario 2, the non-predictable score $\bu_3$ has the highest variance contribution. Thus we expect PCA to identify {\color{black}linear combinations of $\bu_3$ and $\bu_1$ for the first two PCs, with a large contribution of $\bu_3$ for the first PC. Meanwhile,} we anticipate the other predictive methods to still pick {\color{black}linear combinations of} $\bu_1$ and $\bu_2$. 

Table \ref{tab-scen123} shows the results for the prediction R$^2$'s  across 1,000 simulations under scenario 1. As expected under scenario 1, all methods perform comparably when the training data is complete. While the results for MCAR 30\% and 40\% are not shown in this chapter, we observed similar patterns to the three-dimensional simulations where the performance of PCA and PredPCA decreases steadily as the amount of MCAR missing data increases. Under MCAR 35\% setting, ProPrPCA-Spline has the best median R$^2$'s for both PCs. 

Under MAR, data among the first five pollutants are more likely to be missing at locations with extreme geographic covariate values. This setup effectively has an impact on the actual variance contributions of the underlying scores in a given sample, and particularly lowers the variability contributed by $\bu_1$. As a result, for PC1, PCA is likely to produce loadings with higher contribution from $\bu_2$ than before. As the predictive methods (PredPCA and ProPrPCA) attempt to balance out the trade-off between data representativeness and spatial predictability, these methods will also likely to obtain linear combinations with more weights from $\bu_2$ for PC1 than before. Subsequently, linear combinations obtained for PC2 will have more weights from $\bu_1$ than before. This explains the decreases in median R$^2$'s of PC1 for all methods but slight increases for PC2. ProPrPCA-Spline notably has the best median R$^2$ for PC1.

\begin{figure}[H]
	\centering
	\includegraphics[width=6.35in]{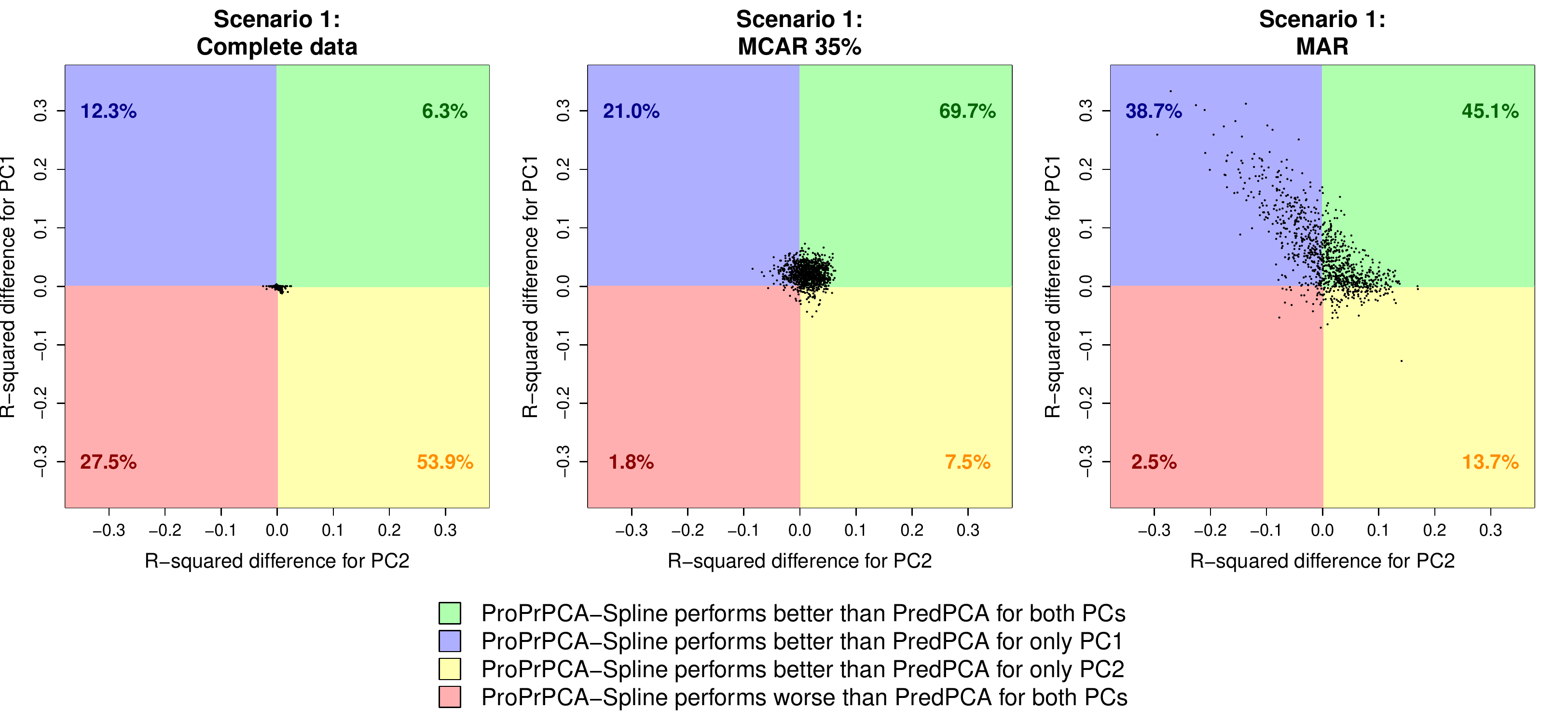}
	\caption{\doublespacing Differences in prediction R$^2$ values between ProPrPCA-Spline and PredPCA for high-dimensional scenario 1. Each dot represents result from one simulation. Percentages indicate the proportion out of 1,000 simulations.} %under complete data, MCAR 35\%, and MAR
	\label{fig-hd-s123}
\end{figure}

We further compare the differences in R$^2$ values between ProPrPCA-Spline and PredPCA in Figure \ref{fig-hd-s123}. With complete training data, ProPrPCA-Spline outperforms PredPCA for only less than 60\% of the simulations, and the magnitude of the difference between the two methods is rather negligible. Under MCAR 35\%, ProPrPCA-Spline outperforms PredPCA for both PCs in 69.7\% of the 1,000 simulations, and, for 28.5\% of the time, ProPrPCA-Spline is better in one of the PCs. Finally, under MAR, there are only 2.5\% of the simulations in which ProPrPCA-Spline is worse than PredPCA for both PCs. There are 38.7\% of the simulations where ProPrPCA-Spline is better for only PC1 (blue top-left quadrant). Particularly for points lying in this quadrant, the greater spread along the y-axis implies that a higher increase in R$^2$ for PC1 is often accompanied by a smaller decrease in  R$^2$ for PC2. Thus ProPrPCA-Spline shows more prominent benefits for PC1 without trading off too much in predictability of PC2. 

\begin{figure}[H]
	\centering
	\includegraphics[width=6.35in]{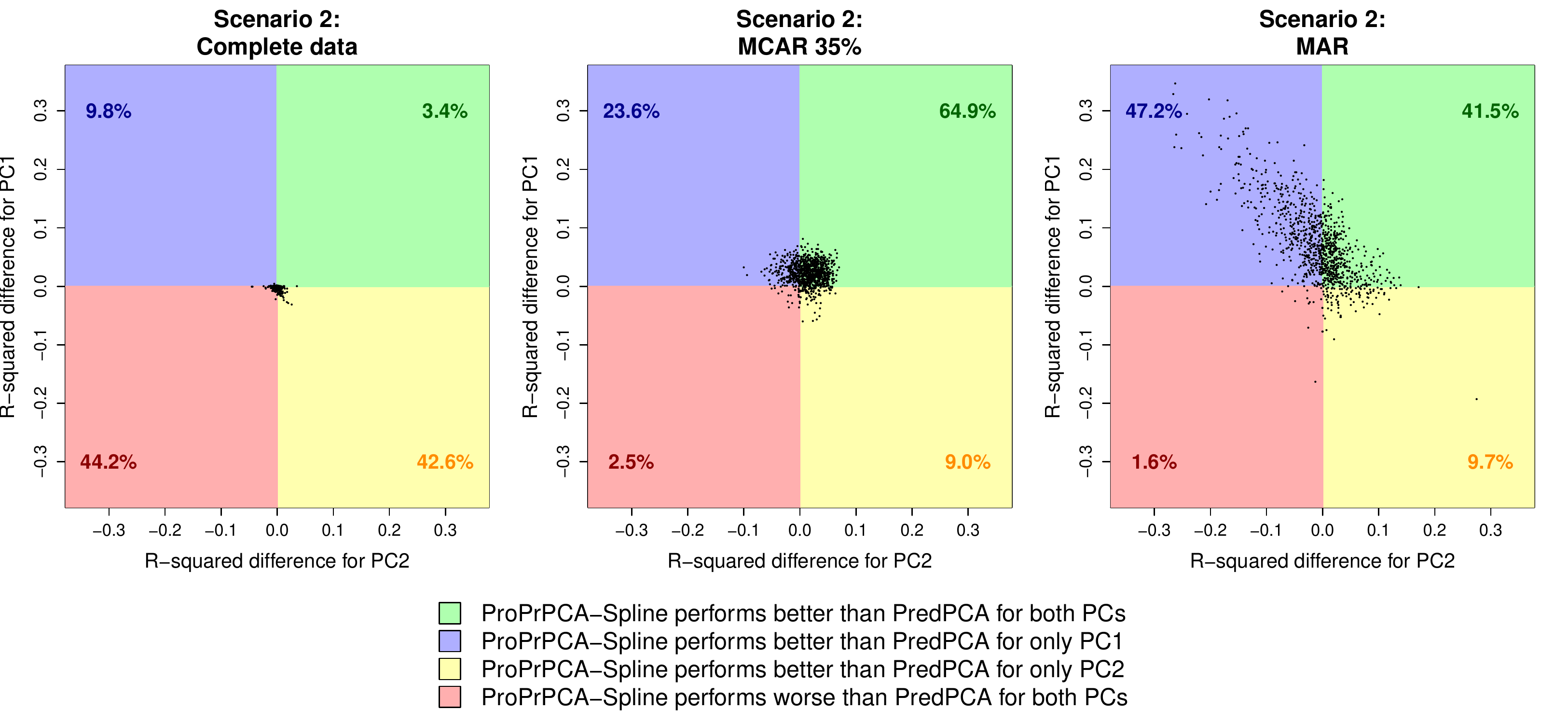}
	\caption{\doublespacing Differences in prediction R$^2$ values between ProPrPCA-Spline and PredPCA for high-dimensional scenario 2. Each dot represents result from one simulation. Percentages indicate the proportion out of 1,000 simulations.} %that fall into each quadrant.
	\label{fig-hd-s312}
\end{figure}

Table \ref{tab-scen312} and Figure \ref{fig-hd-s312} show the corresponding results under scenario 2. In this scenario, as expected, PCA often identifies linear combinations of $\bu_3$ and $\bu_1$ as the first two PCs, and thus the predictive performance is generally poor, especially for  PC1. ProPrPCA-Krige severely underperforms compared to PredPCA and ProPrPCA-Spline, even with complete data. Both PredPCA and ProPrPCA-Spline produce similar median R$^2$'s with complete data. Similar to scenario 1, ProPrPCA-Spline performs consistently well with an increasing amount of MCAR, while the performance of PredPCA deteriorates. ProPrPCA-Spline shows clear benefits under MAR, particularly for PC1 (0.72) compared to PredPCA (0.63). The visualization of the differences in prediction R$^2$'s between ProPrPCA-Spline and PredPCA in Figure \ref{fig-hd-s312} further supports similar conclusions to those of scenario 1. 

\section{Data application \label{sec-application}}

\subsection{Methods}

In this section, we first compare the pollutant profiles obtained by different dimension reduction methods to the annual average 2010 CSN data. Prior to our analysis, we take a similar approach to \cite{keller2017covariate} and convert the mass concentrations of PM$_{2.5}$ components to proportions by dividing by the total mass of PM$_{2.5}$, and then log-transform these proportions. We also follow a similar preprocessing procedure as described in \cite{keller2017covariate} and \cite{jandarov2017novel} to the GIS covariates to be used in the predictive algorithms and spatial prediction model. That is, we remove covariates that are missing at all chosen sites, have the same values in at least 80\% of the sites, or have at least 2\% of their values being more than five standard deviations away from the sample mean. We also remove land-use covariates whose maximal value is only 10\% among all chosen sites. Finally, we apply PCA on the processed GIS data and use the first five PCs in later stages. 

After the preprocessing procedure, we end up with a total of 221 CSN sites, only 130 of which have complete data on all 21 PM$_{2.5}$ components. We first apply three methods, PCA, PredPCA, and ProPrPCA-Spline, on the 130 {\color{black} sites with complete data (the ``complete" set)}. We then proceed to apply these methods on all 221 CSN sites {\color{black} (the ``full" set)}, where LRMC is applied prior to PCA and PredPCA. The goal is to assess how the estimated loadings and PC scores change when using only sites with complete data compared with using all available sites. The design matrix, $\Z$, used in PredPCA and ProPrPCA-Spline includes the five PCs of GIS covariates and thin-plate spline basis functions generated from the spatial coordinates, similar to \cite{jandarov2017novel}. We do not use ProPrPCA-Krige in our comparison because of its computational cost and inferior performance compared to ProPrPCA-Spline in our previously described simulations. 

We also conduct leave-one-site-out cross-validation to compare the predictive performances among these methods. In each round of cross-validation, we leave out one site among the complete sites as test data. We then perform dimension reduction and fit a universal kriging model on training data comprised of either only the remaining complete sites {\color{black} (the ``complete" training data)}, or all remaining sites {\color{black} (the ``full" training data), while the testing data in each round stays the same.} The goal is to assess the predictive performance of different methods with both complete and missing data. 

\subsection{Results}

\subsubsection{The multi-pollutant profile}

\begin{figure}[H]
	\centering
	\includegraphics[width=6.5in]{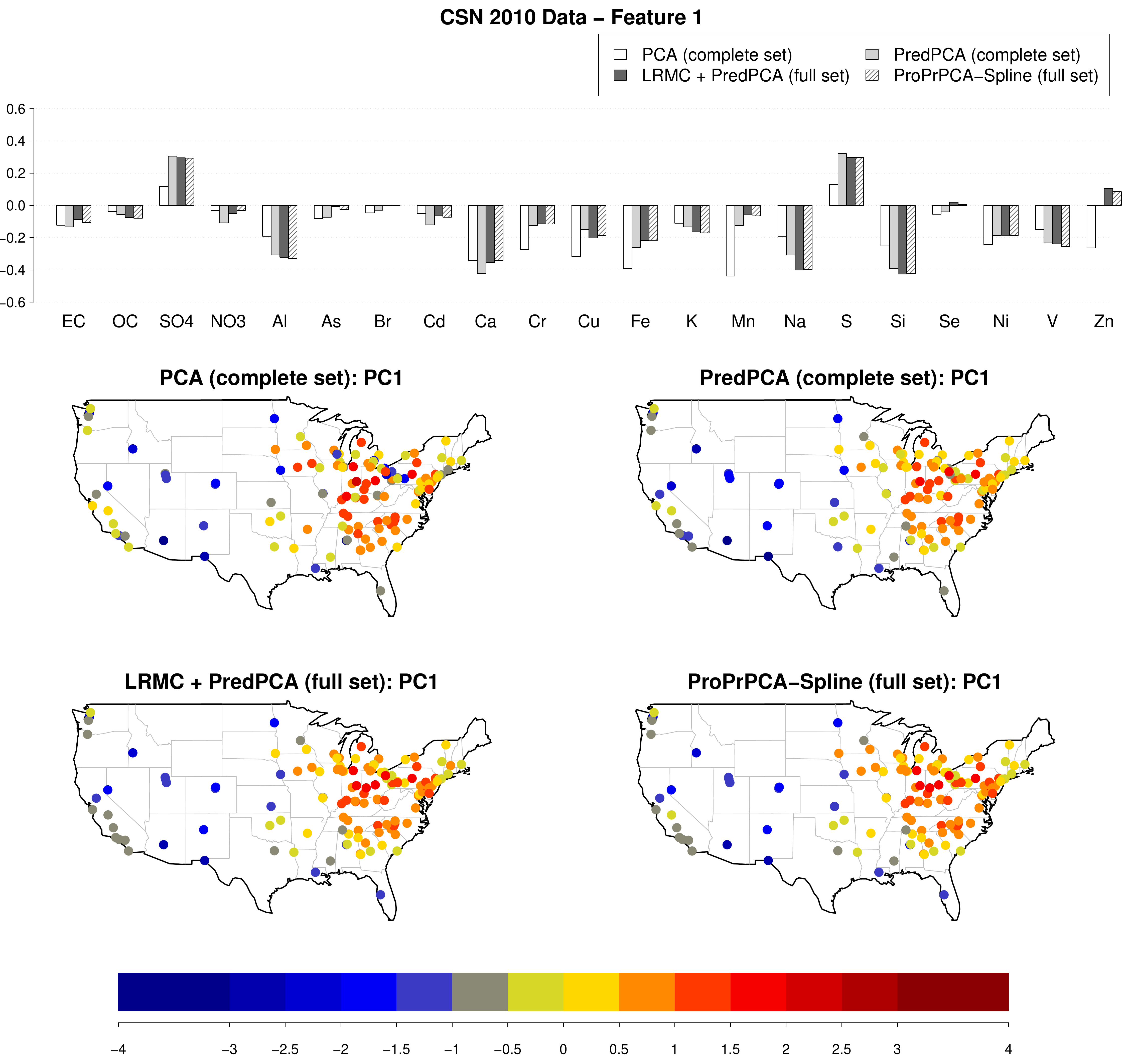}
	\caption{\doublespacing Estimated loadings for feature with highly positive weights on SO$^{2-}_4$ and S, and corresponding scores, obtained from different PCA algorithms applied to 2010 CSN data: PCA and PredPCA applied to the complete set (130 sites with complete data), PredPCA and ProPrPCA-Spline applied to the full set (all 221 available sites).  }
	\label{fig-real-feature1}
\end{figure}

Figure \ref{fig-real-feature1} shows the estimated loadings and the spatial distributions of corresponding scores of the first PC for four combinations of method and dataset: PCA applied to {\color{black} the complete set}, PredPCA applied to {\color{black} the complete set}, imputation followed by PredPCA applied to {\color{black} the full set}, and ProPrPCA-Spline applied to {\color{black} the full set}. The results for ProPrPCA-Spline when using {\color{black} the complete set} (not shown here) are essentially identical to PredPCA results. 

The estimated PC1 loadings are similar across PredPCA applied to either sets and to ProPrPCA-Spline, with highly positive weights on SO$^{2-}_4$ and S and highly negative weights on Al, Ca, Na, and Si. Highly positive scores are observed in the east and part of the Midwest, probably due to sulfur emissions from coal combustion \citep{thurston2011source, hand2012seasonal}. Negative scores are observed in the west and southwest, and have a classic resuspended soil profile \citep{thurston2011source, tong2012long, clements2017source}. While the spatial distribution of PCA scores looks similar to other methods, loadings obtained by PCA applied to {\color{black} the complete set} are fundamentally different than the rest, with much weaker positive weights on SO$^{2-}_4$ and S, and strongly negative weights on many additional elements, including Cr, Cu, Fe, Mn, Ni, Zn.

\begin{figure}[H]
	\centering
	\includegraphics[width=6.5in]{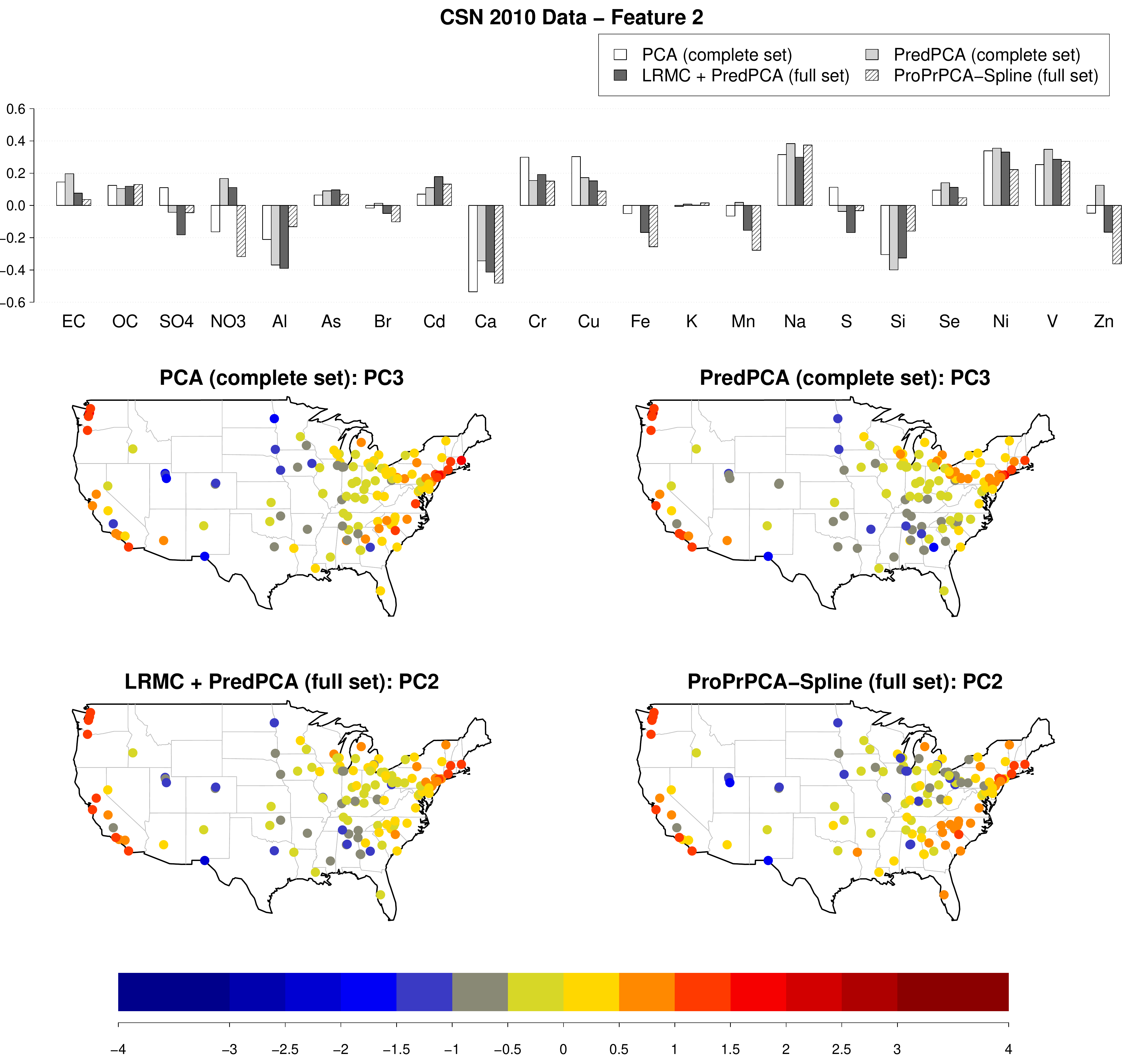}
	\caption{\doublespacing Estimated loadings for feature with highly positive weights on Na, Ni, and V, and corresponding scores, obtained from different PCA algorithms applied to 2010 CSN data: PCA and PredPCA applied to the complete set (130 sites with complete data), PredPCA and ProPrPCA-Spline applied to the full set (all 221 available sites).  }
	\label{fig-real-feature2}
\end{figure}

Figure \ref{fig-real-feature2} shows the estimated loadings and the score distributions for the PC that has a highly positive composition of Na, Ni, and V. This feature corresponds to PC3 obtained by {\color{black}PCA or PredPCA applied to the complete set}, and PC2 obtained by {\color{black}PredPCA or ProPrPCA-Spline applied to the full set}. ProPrPCA-Spline results in highly positive scores along the west coast, the east coast, and southeast region, possibly due to residual oil combustion \citep{thurston2011source}, and marine aerosol \citep{thurston2011source, kotchenruther2017effects}. ProPrPCA-Spline also identifies pronounced negative loadings on Zn and NO$^{-}_3$. The remaining three combinations of methods and datasets are able to produce fairly similar maps with strongly positive scores along the west coast and across the northern east coast, although they fail to highlight some relevant coastal locations in the southeast region.  

\begin{figure}[H]
	\centering
	\includegraphics[width=6.5in]{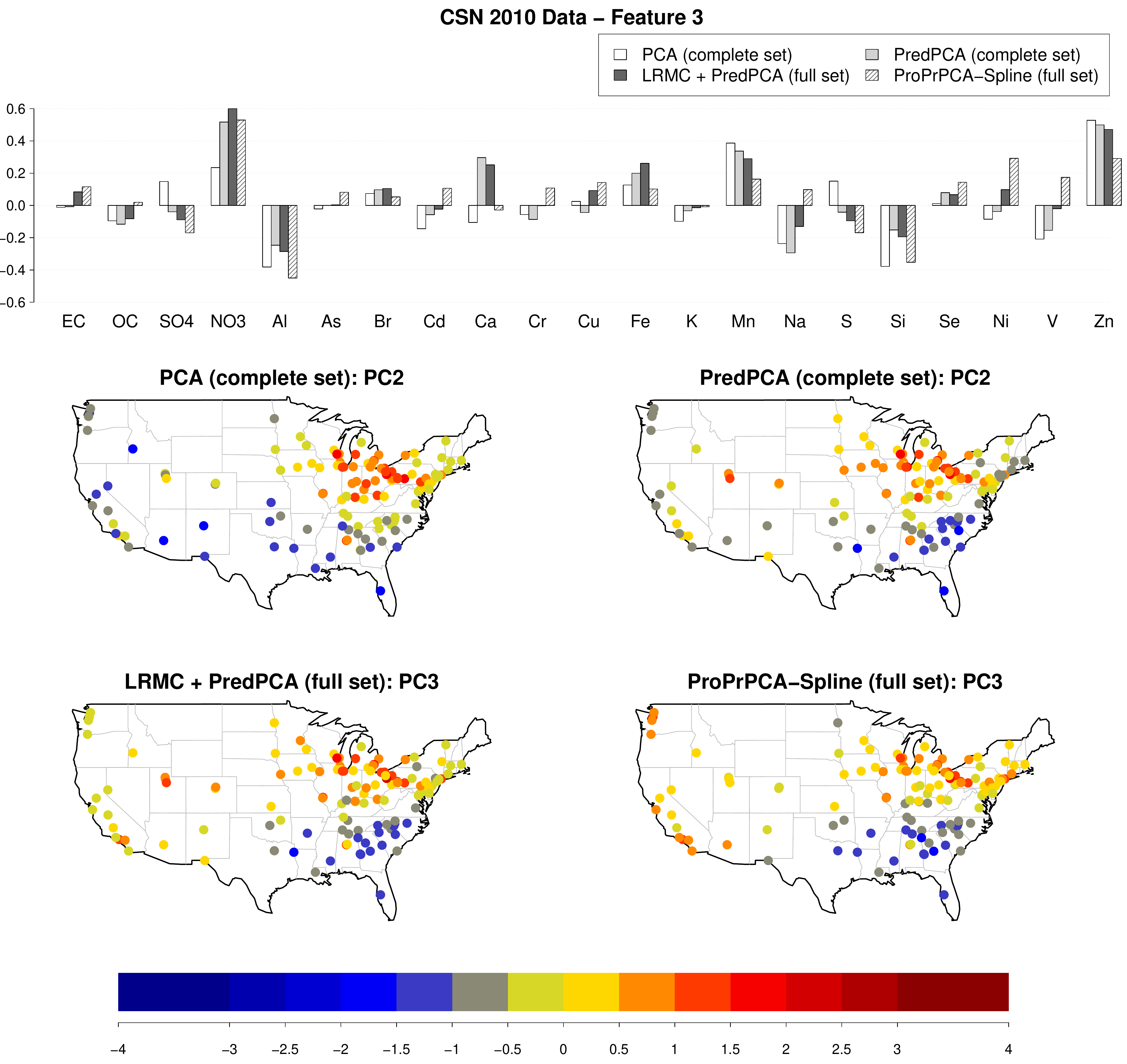}
	\caption{\doublespacing Estimated loadings for feature with highly positive weights on NO$^{-}_3$ and Zn, and corresponding scores, obtained from different PCA algorithms applied to 2010 CSN data: PCA and PredPCA applied to the complete set (130 sites with complete data), PredPCA and ProPrPCA-Spline applied to the full set (all 221 available sites).  }
	\label{fig-real-feature3}
\end{figure}

Figure \ref{fig-real-feature3} shows the results for features highly positive in NO$^{-}_3$ and Zn, which corresponds to PC2 obtained by {\color{black}PCA or PredPCA applied to the complete set}, and PC3 obtained by {\color{black}PredPCA or ProPrPCA-Spline applied to the full set}. For all methods, highly positive scores are observed in the northern Midwest, possibly due to nitrate hazes \citep{coutant2003compilation, pitchford2009characterization, hand2012seasonal}. Additionally, loadings produced by ProPrPCA-Spline are also strongly positive in Ni, V, and negative in Al, Si, with greater magnitude compared to other methods. Thus, moderately positive scores are also observed along the west coast. ProPrPCA-Spline also results in highly positive scores in the southeast region due to the calcium poor soils in that region compared to Al and Si content \citep{shacklette1984element}.

\subsubsection{Cross-validation results}

\begin{figure}[H]
	\centering
	\includegraphics[width=6.5in]{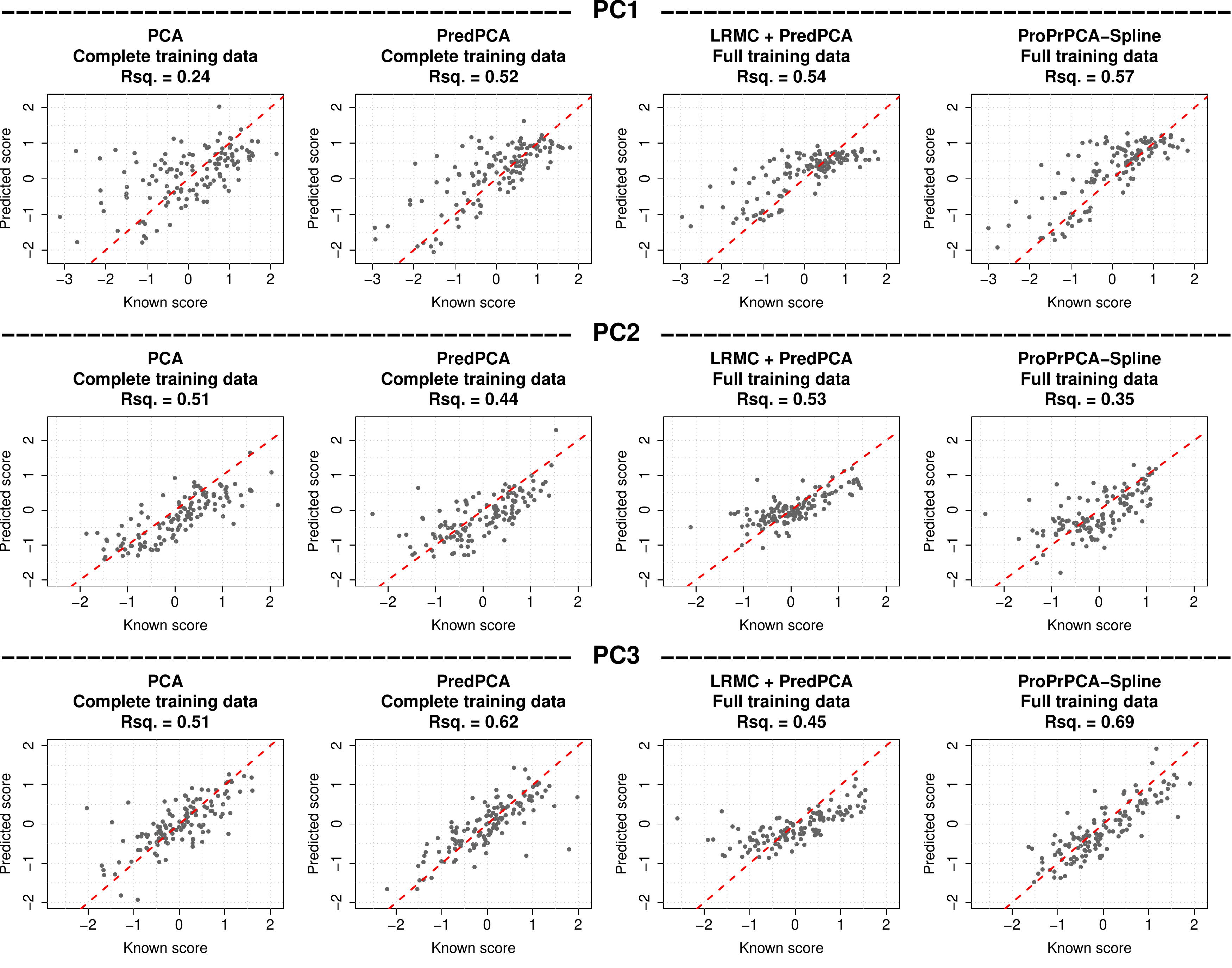}
	\caption{\doublespacing Prediction R$^2$'s from leave-one-site-out cross-validation on 2010 CSN data. Sites with complete PM$_{2.5}$ component data are used as testing data. Training data may include only complete sites, or all available sites.}
	\label{fig-cv}
\end{figure}

Finally, we look at the predictive performances in leave-one-site-out cross-validations, and the results are shown in Figure \ref{fig-cv}. While having decent performance for PC2 and PC3 (R$^2 = 0.51$), using PCA applied to {\color{black} the complete training data} yields a poor result for PC1 (R$^2 = 0.24$). PredPCA has similar performances for PC1 with either {\color{black}complete or full training data}. However, there is a substantial trade-off in performances between PC2 and PC3, which can potentially be explained by the switching between PC2 and PC3 observed in the pollutant profile. ProPrPCA-Spline applied on {\color{black} the full training data} shows the highest predictive performance for PC1 (R$^2 = 0.57$) and PC3 (R$^2 = 0.69$), but suffers from a decrease in the ability to predict PC2 well (R$^2 = 0.35$).

A possible explanation to the overall relatively low R$^2$'s for all methods is that we use the same pre-specified spatial information encoded in $\Z$ to characterize the spatial variability across all PCs, which may not be effective. A potential solution, which is beyond the scope of this paper, is adaptive selection of features to be included in $\Z$, which is proposed and discussed in \cite{bose2018adaptive}. 

\section{Discussion}

In this chapter, we propose a probabilistic extension to the PredPCA algorithm developed by \cite{jandarov2017novel}. The proposed ProPrPCA algorithms can be applied to misaligned multi-pollutant data with missing observations. The ultimate goal is to improve the predictive performance of the exposure modeling stage that is often required in air pollution studies that rely on fixed site monitoring data. {\color{black} In spite of its simplicity, these probabilistic extensions are nontrivial and effective in mitigating the impact of missing data on the predictive performance of the exposure model. The proposed methods also eliminate the necessity of a separate imputation procedure prior to dimension reduction.} The scientific motivation, especially in health-pollution studies on PM$_{2.5}$ and its components, includes the ability to use estimated PC scores at study locations as effect modifiers for the main health associations of interest. 

We have demonstrated via simulations that ProPrPCA-Spline consistently outperforms its competitors under various missing observation scenarios. Its computational speed is on par with both PCA and PredPCA, which are non likelihood-based methods. The complex version, ProPrPCA-Krige, assumes a universal kriging formulation for the latent variable, with the mean model enriched by spatial covariates, and spatial correlations among the residuals. ProPrPCA-Spline incorporates thin-plate spline basis functions, which can be regarded as an alternative to a fixed low-rank kriging model \citep{kammann2003geoadditive}. Intuitively, the latent specification of ProPrPCA-Krige would have been cohesive with the later prediction stage using universal kriging. Possible explanations for the inferior performance of the Krige algorithm in simulations include the difficult nature of the numerical optimization for spatial variance parameters, the number of parameters to estimate, and no guaranteed convergence to the global optima using the EM algorithm. 

PCA is closely related to factor analysis \citep{harman1976modern}, k-mean clustering \citep{macqueen1967some}, or positive matrix factorization \citep{paatero1994positive}, which have recently been used as source apportionment or dimension reduction for exposure data prior to health analyses \citep{sarnat2008fine, ostro2011effects, zanobetti2014health, ljungman2016impact}. These applications, however, have been limited to time-series analysis in specific regions, without the challenge of spatial misalignment and severe missing data. Recent work by \cite{keller2017covariate} and \cite{jandarov2017novel} has modified the traditional clustering and PCA methods, respectively, to the setting of spatially-misaligned multi-pollutant data, where the products of the dimension reduction procedure are desired to be spatially predictable. We further extend these frameworks by considering the realistic challenge of missing monitoring data. Our proposed framework essentially performs model-based imputation, which is cohesive and complementary to the spatial prediction stage. While one can impute the original data with sophisticated low-rank matrix completion techniques, which also operate based on the assumption of a latent variable structure, such methods only rely on observed measures. Therefore, if the missing patterns depend on external geographic covariates, such imputation schemes cannot recover the correct data structure.

In the literature, spatial latent variable models have been explored under the Bayesian framework. For example, \cite{wang2003generalized} proposed a generalized common spatial factor model using MCMC techniques. \cite{hogan2004bayesian} formulated a Bayesian factor analysis model, which was later extended by \cite{liu2005generalized} to motivate a generalized spatial structural equations model, and by \cite{zhu2005generalized} to deal with spatiotemporal data. These rich modeling approaches have not been utilized in the setting of multi-pollutant analysis with spatial misalignment. The main goal of these models is often to explain the associations between the original variables and the underlying factors. Here the goal of an improved PCA algorithm is to obtain a lower-dimensional representation of the data in a spatially predictive way for subsequent use in spatial prediction and health regression.   

The multi-stage procedure in analyzing health-pollution association under spatial misalignment is a common and pragmatic approach \citep{crouse2010postmenopausal, bergen2013national, chan2015long}. However, it is important to be mindful of the potential implications of measurement errors and model uncertainty of the spatial prediction stage on the health inference model, a topic which has been discussed extensively in \cite{szpiro2013measurement}. Additionally, these authors emphasized that the spatially structured components of the covariates used in the health model should be included in the exposure modeling stage to guarantee a consistent estimation of the health effects. In the multi-pollutant setting with missing observations, additional stages of imputation and dimension reduction lead to more complicated layers of uncertainty. Our proposed methods eliminate the need of a separate imputation step prior to dimension reduction, as these two steps are handled simultaneously using a model-based approach. A possible alternative to the multi-stage paradigm is a unified approach where both exposure and health data are considered simultaneously in a joint model, while leveraging the factor analysis framework to perform dimension reduction. \cite{szpiro2013measurement} point out several disadvantages of such joint model, including sensitivity to influential or outlying health data, vulnerability to model mis-specifications, and computational burden, especially with multi-pollutant data. 

While we focus our discussion in this chapter exclusively on studies involving data on PM$_{2.5}$ and its components, our proposed method is both appropriate for other multi-pollutant studies and applicable to other fields in general where spatial misalignment necessitates an exposure modeling procedure. Future work includes further understanding and improvement of the ProPrPCA-Krige algorithm, and a possible extension to spatiotemporal data.

\section{Supporting Information}

Data used in this paper are available upon request through the MESA Air team at the University of Washington. Supplemental materials can be provided upon email request. %Additional information and supporting material for this article is available online at the journal's website. 

\bibliographystyle{apalike}
\bibliography{VuPT-ProPrPCA}

\clearpage

\begin{table}[H]
	\begin{center}
		\caption{ The algorithm for ProPrPCA-Spline with complete monitoring data}
		\label{tab-spline-alg} 
		\begin{tabular}{@{}*{1}{p{\textwidth}@{}}}
			\hline
			\textbf{Input} $\X$, $\Z$, $q$, and $t_{max}$ \\
			\hspace{0.5cm} \textbf{for} $l$ in $\{1, ..., q \}$ \textbf{do} \\
			\hspace{1cm} $\X_l \leftarrow \X_{l-1} - \hat{\bu}_{l-1}\hat{\bv}^\tp_{l-1} $ where $\X_0 = \X$, $\hat{\bu}_0 = \bze$, and $\hat{\bv}_0 = \bze$ \\
			\hspace{1cm} \textbf{Initialize} $\bv^{(0)}_l$, $(\gamma^{(0)}_l)^2$, $\bet^{(0)}_l$, and $t = 1$ \\
			\hspace{1.5cm} \textbf{while} not converged \textbf{or} $t < t_{max}$ \textbf{do} \\
			\hspace{2cm} $\bv^{(t+1)}_l \leftarrow {\tilde{\bv}_l}/{\Vert \tilde{\bv}_l \Vert_2} $ where $\tilde{\bv}_l \leftarrow {\X_l^\tp \Z \bet^{(t)}_l }/{\big\Vert \Z \bet^{(t)}_l \big\Vert^2_2} $   \\
			\hspace{2cm} $\bet^{(t+1)}_l \leftarrow \left( \Z^\tp \Z \right)^{-1} \left(\Z \otimes \bv^{(t+1)}_l \right)^\tp \text{vec}(\X_l) $ \\
			\hspace{2cm} $(\gamma^{(t+1)}_l)^2 \leftarrow  (np)^{-1} \big\Vert \text{vec}(\X_l) - (\I_n \otimes \bv^{(t+1)}_l)\Z \bet^{(t+1)}_l    \big\Vert^2_2$ \\
			\hspace{2cm} $t \leftarrow t+1$ \\
			\hspace{1.5cm} \textbf{end while} \\
			\hspace{1cm} $\hat{\bv}_l \leftarrow \bv^{(t)}_l$, $\hat{\gamma}^2_l \leftarrow (\gamma^{(t)}_l)^2$, $\hat{\bet}_l \leftarrow \bet^{(t)}_l$ \\
			\hspace{1cm} $\hat{\bu}_l = \X_l \hat{\bv}_l$ \\
			\hspace{0.5cm} \textbf{end for} \\
			\textbf{Output} $\{ \hat{\bv}_1, ...,\hat{\bv}_q  \}$, $\{ \hat{\bu}_1, ...,\hat{\bu}_q  \}$, $\{ \hat{\bet}_1, ...,\hat{\bet}_q  \}$, $\{ \hat{\gamma}^2_1, ...,\hat{\gamma}^2_q  \}$ \\
			\hline
		\end{tabular}
	\end{center}
\end{table}

\begin{table}[H]
	\centering
	\caption{\doublespacing Means (standard deviations) of estimated PC1 loadings across 1,000 replications with three-dimensional surface with spatially correlated noise and complete training data.}
	\label{tab-loading-scen2} 
	\begin{tabular}{lccc} 
		\hline
		& $X_1$ &  $X_2$ & $X_3$ \\ \hline
		PCA & 0.40 (0.11) & 0.41 (0.09) & 0.80 (0.07) \\
		PredPCA & 0.88 (0.04) & -0.07 (0.04) & 0.46 (0.09) \\
		ProPrPCA-Krige & 0.85 (0.04) & -0.11 (0.08) & 0.50 (0.08) \\
		ProPrPCA-Spline & 0.86 (0.03) & -0.12 (0.07) & 0.49 (0.07) \\ \hline
	\end{tabular}
\end{table}

\begin{table}[H]
	\caption{\doublespacing Means (standard deviations) of estimated PC1 loadings across 1,000 replications with three-dimensional surface with independent noise and complete training data.}
	\label{tab-loading-scen1} 
	\centering
	\begin{tabular}{lccc}
		\hline
		& $X_1$ &  $X_2$ & $X_3$ \\ \hline
		PCA & 0.53 (0.06) & 0.39 (0.04) & 0.75 (0.03) \\
		PredPCA & 0.89 (0.02) & 0.01 (0.02) & 0.45 (0.04) \\
		ProPrPCA-Krige & 0.88 (0.02) & 0.03 (0.04) & 0.47 (0.04) \\
		ProPrPCA-Spline & 0.89 (0.02) & 0.01 (0.03) & 0.46 (0.04) \\ \hline
	\end{tabular}
\end{table}

\begin{table}[H] 
	\caption{\doublespacing The median prediction R$^2$'s across 1,000 simulations for high-dimensional scenario 1. Under missing data scenarios, LRMC is used prior to either PCA or PredPCA.} 
	\label{tab-scen123} 
	\centering
	{\color{black}
		\begin{tabular}{lccc} 
			\hline
			PC1 & Complete &	MCAR 35\% &	MAR \\ \hline
			PCA & 0.83	& 0.80	& 0.61 \\
			PredPCA & 0.84	&  0.81 & 0.63 \\
			ProPrPCA-Krige & 0.83	& 0.83	& 0.64 \\
			ProPrPCA-Spline & 0.84	& 0.83	& 0.69 \\
			\hline
	\end{tabular}}
	
	\vspace{0.25cm}
	{\color{black}
		\begin{tabular}{lccc}
			\hline
			PC2 & Complete &	MCAR 35\% &	MAR \\ \hline
			PCA & 0.60	& 0.58	& 0.67 \\
			PredPCA & 0.60	& 0.58	& 0.68 \\
			ProPrPCA-Krige & 0.60	& 0.60	& 0.69 \\
			ProPrPCA-Spline & 0.60	& 0.60	& 0.68 \\
			\hline	
	\end{tabular}}
\end{table}  

\begin{table}[H]
	\caption{\doublespacing The median prediction R$^2$'s across 1,000 simulations for high-dimensional scenario 2. Under missing data scenarios, LRMC is used prior to either TradPCA or PredPCA.} 
	\label{tab-scen312} 
	\centering
	\begin{tabular}{lccc}
		\hline
		PC1 & Complete &	MCAR 35\% &	MAR \\ \hline
		PCA & 0.01	& 0.01	& 0.00 \\
		PredPCA & 0.81	& 0.78	& 0.63 \\
		ProPrPCA-Krige & 0.70	& 0.66	& 0.41 \\
		ProPrPCA-Spline & 0.81	& 0.80	& 0.72 \\
		\hline
	\end{tabular}
	
	\vspace{0.25cm}
	\begin{tabular}{lccc}
		\hline
		PC2 & Complete &	MCAR 35\% &	MAR \\ \hline
		PCA & 0.78	& 0.74	& 0.60 \\
		PredPCA & 0.56	& 0.54	& 0.62 \\
		ProPrPCA-Krige & 0.30	& 0.26	& 0.23 \\
		ProPrPCA-Spline & 0.56	& 0.56	& 0.59 \\
		\hline	
	\end{tabular}
\end{table}

\end{document}